\documentclass[12pt]{article}
\usepackage[latin1]{inputenc} 
\usepackage{amssymb}
\usepackage{amsfonts} 
\usepackage{amsthm} 
\usepackage{amsmath} 
\usepackage{hyperref}

\def\nn{\nonumber}

\def \bc {\begin{center}}
\def \ec {\end{center}}
\def \bi {\begin{itemize}}
\def \ei {\end{itemize}}

\def \ba {\begin{array}}
\def \ea {\end{array}}

\def \bea {\begin{eqnarray}}
\def \eea {\end{eqnarray}}

\def \be {\begin{equation}}
\def \ee {\end{equation}}

\def \lb {\left[}
\def \rb {\right]}
\newcommand{\la}{\langle}
\newcommand{\ra}{\rangle}

\def \um {\frac{1}{2}}

\def\tr {\mathrm{tr}}

\def\mbN {{\mathbb N}}

 \def\cD {{\cal D}}

\newtheorem{thm}{Theorem}[section]
\newtheorem{cor}[thm]{Corollary}
\newtheorem{lem}[thm]{Lemma}
\newtheorem{prop}[thm]{Proposition}

\theoremstyle{remark}
\newtheorem{rem}[thm]{Remark}

\begin{document}

\begin{center}
{\Large {\bf On the oscillator realization of conformal $U(2,2)$ quantum 
particles and their particle-hole coherent states}}
\end{center}
\bigskip

\centerline{{\sc M. Calixto} 
 and {\sc E. Pérez-Romero}}

\bigskip

\bc {\it  Departamento de Matemática Aplicada,
Facultad de Ciencias, Campus de Fuentenueva,  18071 Granada, Spain}
\ec

\bigskip
\begin{center}
{\bf Abstract}
\end{center}
\small
\begin{list}{}{\setlength{\leftmargin}{3pc}\setlength{\rightmargin}{3pc}}
\item
We revise the unireps. of $U(2,2)$ describing conformal particles with continuous mass spectrum from a many-body perspective, which 
shows massive conformal particles as compounds of two correlated massless particles. The statistics of the compound 
(boson/fermion) depends on the helicity $h$ of the massless components (integer/half-integer). Coherent states (CS) of particle-hole 
pairs (``excitons'') are also explicitly constructed as the exponential action of exciton (non-canonical) creation operators on the 
ground state of unpaired particles. These CS are labeled by points $Z$ ($2\times 2$ complex matrices) on the Cartan-Bergman domain 
$\mathbb D_4=U(2,2)/U(2)^2$, and constitute a generalized (matrix) version of Perelomov $U(1,1)$ coherent states labeled by points 
$z$ on the unit disk $\mathbb D_1=U(1,1)/U(1)^2$. Firstly we follow a geometric approach to the construction of CS, orthonormal 
basis, $U(2,2)$ generators and their matrix elements and symbols in the reproducing kernel Hilbert space 
$\mathcal H_\lambda(\mathbb D_4)$ of analytic square-integrable holomorphic functions on $\mathbb D_4$, 
which carries a unitary irreducible representation of $U(2,2)$ with index $\lambda\in\mathbb N$ (the conformal or scale dimension). Then we introduce 
a many-body representation of the previous construction through an oscillator realization of the $U(2,2)$ Lie algebra generators 
in terms of eight boson operators with constraints. 
This particle picture allows us for a physical interpretation of our abstract mathematical construction in the 
many-body jargon. In particular, the index $\lambda$ is related to the number $2(\lambda-2)$ of unpaired quanta  and to the 
helicity $h=(\lambda-2)/2$ of each massless particle forming the massive compound.
\end{list}
\normalsize 




\section{Introduction}

The generalization of Poincaré symmetry of space-time to conformal symmetry is a recurrent subject with a vast literature 
in mathematical and particle physics. However, its physical interpretation and its broken character keep raising 
controversy. Special conformal transformations can be either interpreted as transitions to systems of relativistic, 
uniformly accelerated observers (see e.g. Refs.\cite{Hill,Cervero1975,conforme-ac}), the temporal component 
being a kind of kinematical red-shift (resembling Hubble's law) \cite{Hill2}, or related to the
Weyl's idea of different lengths in different points of space time
\cite{Weyl}, or to the Kastrup's interpretation as geometrical gauge transformations of the Minkowski space \cite{Kastrup1}, etc. 
Moreover, it is usually assumed that exact scale invariance is physically unacceptable since it implies that the mass 
spectrum is either continuous or all masses are zero.  

In this article we revisit the conformal group $SU(2,2)$ and some of its positive mass unirreps (discrete series, 
to be more precise) from an oscillator realization that provides an interesting many-body interpretation. 
Since Jordan \cite{Jordan} (see also \cite{Louck3}) and Schwinger \cite{Schwingerbook} introduced a way of representing the matrix generators
of a certain symmetry in terms of bilinear products (field operators) of either boson or fermion
type, this oscillator representation (also called Jordan-Schwinger mapping) has been widely used to provide a treatment of representations
of Lie groups, with the  calculation of matrix elements of finite and infinitesimal group transformations 
in the bases of coherent and Fock states; see, e.g., Biedenharn \& Louck \cite{Louck3} for unitary groups, Moshinsky et al. 
\cite{MoshinskyPL,MoshinskyNPB,MoshinskyJMP,MoshinskyBook} and Iachello et al. \cite{book1,book2,book3} 
in the context of molecular, atomic and nuclear physics,  and \cite{Manko1,Manko2,Manko3} for the case 
of Cayley-Klein groups. Many-body quantum systems in the second quantized field formalism 
fit well into this picture, where Hamiltonians (and other relevant operators) are multidimensional quadratic 
in boson or fermion creation and annihilation operators. 

For the conformal group $SU(2,2)$, the oscillator realization of the so called \emph{ladder representations}, describing massless particles, where studied long time ago by 
\cite{Mack-Todorov,Todorov}.  This Jordan-Schwinger-like mapping is in fact extensible to general pseudo-unitary grups $U(p,q)$ \cite{Todorov2,couplingUpq} and it became 
popular after \cite{u66}, who discussed the use of $U(6,6)$ to classify hadrons; in this case barions and antibarions belong to mutually conjugate representations 
with respect to $U(6)$. The case of positive-mass unireps. of $U(2,2)$ was already discussed in, for example, \cite{Mack,Ruhl0,Ruhl1}, 
and more recently by us in \cite{EMSMTA,spinning,unruh},  but their oscillator realizations have not been studied as thoroughly as the simpler case of 
ladder (most degenerate) representations. An inspiring article on this subject has been recently published in \cite{Grosse}, in the context of deformation quantization. 
Here we further develop this oscillator realization of discrete series representations of $SU(2,2)$ by providing explicit expressions for orthonormal basis vectors, 
coherent states (see standard references and  textbooks \cite{Klauder,Perelomovbook,Perelomov,Gazeau,Gazeaubook} on this subject), operators and their matrix elements and symbols. 
As far as we know, explicit expressions of this kind have not been written before and we 
think that they will be useful not only for the better understanding of the structure of conformal quantum particles 
but also in some condensed matter applications related to pairing (particle-hole) models (see later in Section \ref{comments} for a physical interpretation 
of our construction in this sense).

The oscillator realization of $U(2,2)$ fits well into the \emph{twistor program} introduced by R. Penrose and coworkers in the 1960's 
\cite{Penrose1,Penrose2,Penrose-MacCallum,Penrose3,Penrose-Rindler} as an approach 
to the unification of quantum theory with gravity.  These constructions had wide application across pure 
and applied mathematics, but not so extensive in basic physics, mainly because twistor space is chiral and treats 
the left and right handed parts of physical fields differently. However, Witten's  paper \cite{Witten} on 
twistorial representations of scattering amplitudes showed how to overcome this, and left-right symmetric theories on space-time 
naturally arise when string theory is introduced into twistor space. This is the subject of much recent activity, with a reworking 
on the theory of particle interactions in twistor language (see e.g. \cite{Hodges} and references therein). We comment in Sec. \ref{comments} 
on this twistor picture and its relation to the oscillator realization by making use of nonlinear sigma-model Lagrangians 
on cosets $\mathbb D_3=U(2,2)/U(2,1)\times U(1)$ (for massless particles) and $\mathbb D_4=U(2,2)/U(2)^2$ (for massive particles), the latter 
being related to the forward tube domain of the complex Minkowski space. 

The paper is organized as follows. In Section \ref{su4coord} we briefly remind the Lie algebra structure and coordinate systems 
of $U(2,2)$ adapted to the fibration $U(2)^2\to U(2,2)\to \mathbb D_4$. In Section \ref{sec2} we construct a coherent state (CS) 
system labeled by points of $\mathbb D_4$ in the (reproducing kernel) Hilbert space ${\cal H}_\lambda(\mathbb D_4)$ 
of analytic square-integrable holomorphic functions on $\mathbb D_4$ with a given measure. This corresponds to a given  
square-integrable irreducible representation of $U(2,2)$ with positive integer index $\lambda$ representing the conformal or scale 
dimension. Firstly we follow a geometric approach to the construction of CS on $\mathbb D_4$, in part inspired by the method of orbits in geometric quantization  
due to Kirillov-Kostant-Souriau \cite{Kirillov,Kostant,Souriau} and the Borel-Weil-Bott theorem \cite{FultonHarris}, 
which relate  quantization, geometry and the representation theory
for classical groups. In Section \ref{sec3} we explicitly compute the infinitesimal generators (Poincaré plus dilations and special conformal transformations) 
of the representation of $U(2,2)$ on ${\cal H}_\lambda(\mathbb D_4)$, their matrix elements in an orthonormal basis and their operator symbols 
in coherent states. Then in Section \ref{oscisec} we introduce an oscillator realization of the 
$u(2,2)$ Lie algebra in terms of eight boson creation, $a_\mu^\dag, b_\mu^\dag$, and annihilation,
$a_\mu, b_\mu, \mu=0,1,2,3$, operators, and express the orthonormal basis of ${\cal H}_\lambda(\mathbb D_4)$ 
in terms of the Fock basis with constraints on the occupancy numbers.  An expression of $\mathbb D_4$ CS as 
Bose-Einstein-like condensates of excitons is also provided. This way we connect the abstract construction of Section \ref{sec2} 
with the ``many-body picture''. This realization differs from the standard boson representation of $u(2,2)$ in terms of 
four  bosons, leading to ladder representations for massless particles and related to the pseudo complex projective $\mathbb D_3$. 
Actually, conformal massive particles turn out to be a compound of two massless correlated particles, something that might 
result strange in principle. The compound wavefunctions are symmetric (resp. antisymmetric) under the interchange of the two massless 
constituents for $\lambda$ even (resp. odd), the helicity of each massless particle being $(\lambda-2)/2$. 
The 0+1 dimensional case, which is described by Perelomov $su(1,1)$ CS on the unit disk  
$\mathbb{D}_1=U(1,1)/U(1)^2$, is treated in parallel all along the paper, to better appreciate the role played by spin and to 
stress the similarities and differences between  $\mathbb D_4$ and $\mathbb D_1$ CS, the former  being a generalized (matrix) version of the latter. 
Section \ref{comments} is devoted to some comments on the Lagrangian picture and possible physical interpretations, not only for the better 
understanding of the structure of conformal quantum particles, but also for possible applications in 
general pairing systems like nuclear pairing phenomenon, superconductivity in solid state physics, Bose-Einstein condensation of excitons, etc.

\section{\label{su4coord}$U(2,2)$ coordinate systems and generators}

The usual action of the conformal group $SO(4,2)$ on Minkowski
spacetime $\mathbb R^4\ni x^\mu$ is comprised of Poincar\'e spacetime
translations $x'^\mu = x^\mu+b^\mu$ and  Lorentz  transformations $x'^\mu=\Lambda^{\mu}_\nu x^\nu$ augmented by dilations  
$x'^\mu=\rho x^\mu$  and relativistic uniform accelerations (special conformal
transformations) $x'^\mu=\frac{x^\mu+a^\mu x^2}{1+2a x+a^2
x^2}$. We shall denote by $P^\mu, M^{\mu\nu}, D$ and $K^\mu$ the corresponding Lie algebra generators, respectively. 
Here we are interested in the usual $4\times 4$ matrix realization of these conformal Lie algebra generators 
\be\ba{rcl} D&=&\frac{\gamma^5}{2},\;M^{\mu\nu}=\frac{\lb
\gamma^\mu,\gamma^\nu\rb}{4}=\frac{1}{4}\left(\ba{cc} \sigma^\mu\check{\sigma}^\nu-\sigma^\nu\check{\sigma}^\mu & 0\\
0&\check{\sigma}^\mu\sigma^\nu-\check{\sigma}^\nu\sigma^\mu\ea\right),\\
P^\mu&=&\gamma^\mu\frac{1+\gamma^5}{2}=\left(\ba{cc} 0& \sigma^\mu \\ 0
&0\ea\right),\;K^\mu=\gamma^\mu\frac{1-\gamma^5}{2}=\left(\ba{cc} 0& 0
\\ \check{\sigma}^\mu &0\ea\right)\ea\label{confalgamma}\ee
in terms of gamma matrices $\gamma^\mu$ in, for instance, the Weyl basis
\be \gamma^\mu=\left(\ba{cc} 0& \sigma^\mu \\ \check{\sigma}^\mu
&0\ea\right),\;\;
\gamma^5=i\gamma^0\gamma^1\gamma^2\gamma^3=\left(\ba{cc}
-\sigma^0& 0\\ 0& \sigma^0\ea\right),\nn\ee
where $\check{\sigma}^\mu\equiv \sigma_\mu=\eta_{\mu\nu}\sigma^\nu$ [we are using the
convention $\eta={\rm diag}(1,-1,-1,-1)$ for the Minkowski metric] 
and $\sigma^\mu$ are the Pauli matrices (plus identity $\sigma^0$)
\be \sigma^0=\left(\ba{cc} 1& 0
\\ 0 &1\ea\right),\;\sigma^1=\left(\ba{cc} 0& 1
\\ 1 &0\ea\right),\;\sigma^2=\left(\ba{cc} 0& -i
\\ i &0\ea\right),\;\sigma^3=\left(\ba{cc} 1& 0
\\ 0 &-1\ea\right).\nn\ee
These are the Lie algebra generators of the fundamental representation
of the four cover of $SO(4,2)$:
\be SU(2,2)=\left\{g=\left(\ba{cc} A& B
\\ C &D\ea\right)\in {\rm Mat}_{4\times 4}(\mathbb C):  g^\dag \Gamma g=\Gamma,
\det(g)=1\right\},\label{su22} \ee
with $\Gamma$ a ${4\times 4}$ hermitian form of signature
$(++--)$. In particular, taking $\Gamma=\gamma^5$, the $2\times 2$
complex matrices $A,B,C,D$ in (\ref{su22}) satisfy the following
restrictions:
\be g^{-1}g=I_{4}\Leftrightarrow \left\{\ba{r} D^\dag
D-B^\dag B=\sigma^0,\\ A^\dag A-C^\dag C=\sigma^0,\\ A^\dag B-C^\dag
D=0,\ea\right.\label{mim}\ee
together with those of $gg^{-1}=I_{4}$. In this article we
shall work with $SU(2,2)$ instead of $SO(4,2)$ and we shall use a
set of complex coordinates to parametrize $SU(2,2)$. This parametrization will be adapted
to the non-compact complex Grassmannian $\mathbb D_4=U(2,2)/U(2)^2$ of the
maximal compact subgroup $U(2)^2=U(2)\times U(2)$. It can be obtained through a
block-orthonormalization process with metric $\Gamma=\gamma^5$ of
the matrix columns of:
\be \left(\ba{cc} \sigma^0& 0
\\ Z^\dag &\sigma^0\ea\right)\rightarrow g=\left(\ba{cc} \Delta_1& Z\Delta_2
\\ Z^\dag\Delta_1 &\Delta_2\ea\right), \left\{ \ba{l} \Delta_1=(\sigma^0-ZZ^\dag)^{-1/2},\\
\\ \Delta_2=(\sigma^0-Z^\dag Z)^{-1/2}.\ea\right.
\label{blockortho}\ee
Actually, we can identify
\be Z=Z(g)=BD^{-1}, Z^\dag=Z^\dag(g)=CA^{-1},
\Delta_1=(AA^\dag)^{1/2},\Delta_2=(DD^\dag)^{1/2}\label{zeta}.\ee
From (\ref{mim}), we obtain the positive-matrix conditions
$AA^\dag>0$ and $DD^\dag>0$, which are equivalent to:
\be \sigma^0-ZZ^\dag>0,\; \sigma^0-Z^\dag Z>0,\label{positive}\ee
and define the eight-dimensional symmetric complex
Cartan-Bergman domain
\be \mathbb D_4=U(2,2)/U(2)^2=\{Z\in {\rm Mat}_{2\times 2}(\mathbb C):
\sigma^0-ZZ^\dag>0\} \label{cartandomain}\ee
Moreover, the compactified Minkowski space $\mathbb M=\mathbb S^3\times_{\mathbb Z_2} \mathbb
S^1$ is precisely the Shilov boundary $U(2)=\{Z\in {\rm Mat}_{2\times 2}(\mathbb
C): Z^\dag Z=ZZ^\dag=\sigma^0\}$ of $\mathbb D_4$.

There is a one-to-one mapping from $\mathbb D_4$ onto the future
tube domain
\be \mathbb T_4=\{W=X+iY\in {\rm Mat}_{2\times 2}(\mathbb C):\,
Y>0\},\label{tube}\ee
of the complex Minkowski space $\mathbb C^{4}$, with
$X=x_\mu\sigma^\mu$ and $Y=y_\mu\sigma^\mu$ hermitian matrices and
$Y>0\Leftrightarrow y_0>\|\vec{y}\|$. The (phase space) coordinates $x_\mu$ and $y_\mu$ are related to four-position and four-momenta, respectively. 
This map is given by the
Cayley transformation and its inverse:
\be Z\to W(Z)=i(\sigma^0-Z)(\sigma^0+Z)^{-1},\;\;W\to
Z(W)=(\sigma^0-iW)^{-1}(\sigma^0+iW).\label{Cayley}\ee
This is the 3+1-dimensional analogue of the usual map from the
unit disk onto the upper half-plane in two dimensions. Actually,
the forward tube domain $\mathbb T_4$ is naturally homeomorphic to
the quotient $U(2,2)/U(2)^2$ in a new realization of $U(2,2)$ in terms of
complex $4\times 4$ matrices $f$ which preserve $\Gamma=\gamma^0$ (that is, $f^\dag\gamma^0 f=\gamma^0$), 
instead of matrices $g$ preserving 
$\Gamma=\gamma^5$ and fulfilling \eqref{mim}. Both
realizations of $U(2,2)$ are related by the map
\be g\to f=\Upsilon g \Upsilon^{-1}, \;\;\Upsilon=\frac{1}{\sqrt{2}}\left(\ba{cc} \sigma^0& -\sigma^0\\
\sigma^0& \sigma^0 \ea\right).\label{upsilon}\ee
In this article we shall mainly work with the $\Gamma=\gamma^5$ realization (the interested reader can see more 
details about the tube domain realization in, for example, \cite{spinning} and \cite{EMSMTA}). 

Let us proceed by giving a complete local parametrization of $U(2,2)$ adapted
to the fibration $U(2)^2\to U(2,2)\to \mathbb D_4$. Any element $g\in G$ (in the
present patch, containing the identity element) admits the Iwasawa
decomposition
\be g=\left(\ba{cc} A& B
\\ C &D\ea\right)=\left(\ba{cc} \Delta_1& Z\Delta_2
\\ Z^\dag\Delta_1 &\Delta_2\ea\right)\left(\ba{cc} U_1& 0
\\ 0 &U_2\ea\right),\label{Iwasawa}\ee
where the last factor $U_1=\Delta_1^{-1}A$ and  $U_2=\Delta_2^{-1}D$ 
belongs to $U(2)^2$; i.e., $U_1,U_2\in U(2)$. Likewise, a
parametrization of any $U\in U(2)$ (in a patch containing the
identity), adapted to the quotient $\mathbb S^2=U(2)/U(1)^2$, is
(the Hopf fibration)
\be U=\left(\ba{cc} \mathrm a&  \mathrm b
\\  \mathrm c &  \mathrm d\ea\right)=\left(\ba{cc} \delta & z\delta
\\ -\bar{z}\delta & \delta\ea\right)\left(\ba{cc} u_1& 0
\\ 0 & u_2\ea\right),\label{Iwasawa2}\ee
where $z= \mathrm b/ \mathrm d\in \overline{\mathbb C}\simeq \mathbb S^2$ (the
one-point compactification of $\mathbb C$ by inverse stereographic
projection), $\delta=(1+z\bar{z})^{-1/2}$ and the phases $u_1= \mathrm a/| \mathrm a|,
u_2= \mathrm d/| \mathrm d|$.

\section{Coherent states on $U(2,2)/U(2)^2$, closure relations and orthonormal basis}\label{sec2}

Firstly, let us consider the Hilbert space $L^2(U(2,2),d\mu)$ of square integrable complex
functions $\psi(g)$ on $U(2,2)$ with  invariant
scalar product
\be\la \psi|\psi'\ra=\int_{U(2,2)}
d\mu(g){\psi(g)}\overline{\psi'(g)}\label{scalarprod}\ee
given through the invariant Haar measure $d\mu(g)$, which can be decomposed as:
\be\ba{rcl}
d\mu(g)&=&
\left.d\mu(g)\right|_{\mathbb D_4}\left.d\mu(g)\right|_{U(2)^2},\\
\left.d\mu(g)\right|_{\mathbb D_4}&=&
\det(\sigma_0-Z^\dag Z)^{-4}|dZ|,\\
\left.d\mu(g)\right|_{U(2)^2}&=& dv(U_1)dv(U_2),\ea\label{Haarmeasure}\ee
where we are denoting by $dv(U)$ the Haar measure on $U(2)$,
which can be in turn decomposed as:
\bea
dv(U)&=& \left.dv(U)\right|_{\mathbb S^2} \left.dv(U)\right|_{U(1)^2},\nn\\
\left.dv(U)\right|_{\mathbb S^2}
 &=& (1+z\bar
z)^{-2}|dz|,\label{haarmeasures2}\\
\left.dv(U)\right|_{U(1)^2}&=& -\bar{u}_1du_1 \bar{u}_2du_2.\nn \eea
We have used the Iwasawa decomposition of an element $g$ given in
(\ref{Iwasawa},\ref{Iwasawa2}) and denoted by $|dz|$ and $|dZ|$
the Lebesgue measures on  $\mathbb C$ and $\mathbb C^4$,
respectively (see \cite{EMSMTA} for more explicit expressions of this measure).
The group $U(2,2)$ is represented on $L^2(U(2,2),d\mu)$ as (left-action) $[\mathcal{U}(g')\psi](g)=\psi(g'^{-1}g)$. This representation is
reducible and we shall restrict it to an irreducible subspace. As we want to restrict ourselves
to the quotient $U(2,2)/U(2)^2$, we chose as fiducial  (ground state, lowest weight) vector
$\psi_0^\lambda(g)\equiv\det(D)^{-\lambda}$ for $g$ given in \eqref{Iwasawa} and $\lambda$ an integer number that will eventually
label the corresponding irreducible representation (the so called ``scale dimension'' \cite{Ruhl0}, which is also related to 
the helicity as we comment later in Sec. \ref{oscisec}). In fact,
$\psi_0^\lambda(g)$ is invariant (up to a phase) under $U(2)^2\subset U(2,2)$ since, for $g'=\left(\ba{cc} U_1& 0
\\ 0 &U_2\ea\right)\in U(2)^2$, we have
\be
\psi_0^\lambda(g'^{-1}g)=\det(U_2^\dag D)^{-\lambda}=\det(U_2^\dag)^{-\lambda}\psi_0^\lambda(g).
\ee
Under  a general element $g'=\left(\ba{cc} A'& B'
\\ C' & D'\ea\right)\in U(4)$, the vector $\psi_0^\lambda$ transforms as
\be
\psi_{g'}^\lambda(g)\equiv \psi_0^\lambda(g'^{-1}g)=\det(D'^\dag D-B'^\dag B)^\lambda=
\det(D'^\dag-B'^\dag Z)^\lambda\psi_0^\lambda(g)
,\label{cs}
\ee
where we have used the relations \eqref{zeta} to write $Z=BD^{-1}$. The set of functions in the orbit of $\psi^\lambda_0$ under $U(2,2)$
\be
\mathcal{S}_\lambda=\{\psi_{g}^\lambda\equiv\mathcal{U}(g)\psi^\lambda_0,\; g\in U(2,2)\}
\ee
defines a system of CS.  Note that $\psi_{g}^\lambda$ and $\psi_{g'}^\lambda$ are equivalent (up to a phase) if
$g'g^\dag\in U(2)^2\subset U(2,2)$. We shall prove that this coherent state system  fulfills the resolution of the identity
\be
1=c_\lambda\int_{\mathbb D_4} \left.d\mu(g)\right|_{\mathbb D_4} |\psi^\lambda_g\rangle \langle \psi^\lambda_g|,\label{resol}
\ee
with a suitable normalization constant $c_\lambda$. Before, let us briefly review some auxiliary results.
Note that,  introducing  $Z'^\dag=D'^{\dag-1}B'^\dag$ as in \eqref{zeta},  the state \eqref{cs} can be written as
\be
\psi_{g'}^\lambda(g)=\det(\sigma_0-Z'^\dag Z)^\lambda\overline{\psi_0^\lambda(g')}\psi_0^\lambda(g).\label{cs2}
\ee
We also realize that $|\psi_0^\lambda(g)|^2=\det(DD^\dag)^{-\lambda}=\det(\sigma_0-Z^\dag Z)^{-\lambda}$. To prove \eqref{resol}, we would like
to have before an expansion of $\det(\sigma_0-Z'^\dag Z)^\lambda$ in terms of orthogonal polynomials. The following identity was proved in 
\cite{EMSMTA}.

{\lem \label{antikernelprop} Let  us denote by
\bea
\cD^{j}_{q_a,q_b}(X)=\sqrt{\frac{(j+q_a)!(j-q_a)!}{(j+q_b)!(j-q_b)!}}
\sum_{k=\max(0,q_a+q_b)}^{\min(j+q_a,j+q_b)}
\binom{j+q_b}{k}\binom{j-q_b}{k-q_a-q_b}\nn\\ \times  x_{11}^k
x_{12}^{j+q_a-k}x_{21}^{j+q_b-k}x_{22}^{k-q_a-q_b},\label{Wignerf}\eea
the usual Wigner's $\cD$-matrices for $SU(2)$ (see e.g. \cite{Louck3}), where $j\in
\mbN/2$ (the spin) runs on all non-negative half-integers and
$q_a,q_b=-j,-j+1,\dots,j-1,j$, and $X$ represents here an arbitrary  $2\times 2$
complex matrix with entries $x_{uv}$. For every $
\lambda  \in \mathbb{N}$, $\lambda\geq 2$,  the following identity holds:
\bea
 \sum_{j\in\mbN/2}\frac{2j+1}{\lambda-1}\sum^{\infty}_{m=0}t^{2j+2m}
\binom{m+\lambda-2}{\lambda-2}\binom{m+2j+\lambda -1}{\lambda-2}
 \det(X)^{m}\sum^{j}_{q=-j}\cD^{j}_{qq}(X)\nn\\ ={\det(\sigma^0-tX)^{-\lambda }}.\label{antikernel}\eea
where the sum on $j$ runs over all half-nonnegative integers:  $j=0,\um,1,\frac{3}{2},2,\dots$.
}

\noindent From this Lemma, the following interesting result can be easily proved.

{\thm\label{GMSMT}  The infinite set of homogeneous polynomials
\be
\varphi_{q_1,q_2}^{j,m}(Z)=\sqrt{\frac{2j+1}{\lambda-1}\binom{m+\lambda-2}{\lambda-2}\binom{m+2j+\lambda
-1}{\lambda-2}}\det(Z)^{m}\cD^{j}_{q_1,q_2}(Z),\label{basisfunc}\ee
of degree $2j+2m$ verifies the following
closure relation (the reproducing Bergman kernel):
\be\sum_{j\in\mbN/2}\sum^{\infty}_{m=0}\sum^{j}_{q_1,q_2=-j}
\overline{\varphi_{q_1,q_2}^{j,m}({Z})}\varphi_{q_1,q_2}^{j,m}(Z')=\frac{1}{\det(\sigma^0-Z^\dag
Z')^\lambda}\label{closure}\ee
and constitutes an orthonormal basis of Hilbert space ${\cal H}_\lambda(\mathbb
D_4)=L^2_h(\mathbb D_4,d\mu_\lambda)$ of analytic square-integrable holomorphic functions on $\mathbb D_4$ with measure
\be
d\mu_\lambda(Z,Z^\dag)\equiv c_\lambda|\psi_0^\lambda(g)|^2\left.d\mu(g)\right|_{\mathbb D_4}=
c_\lambda\det(\sigma_0-Z^\dag Z)^{\lambda-4}|dZ|,
\label{projintmeasure}
\ee
where $c_\lambda= {(\lambda-1)(\lambda-2)^2(\lambda-3)}/{\pi^4}$ is a normalization constant and $\lambda>3$.}

\noindent This theorem has been proved in \cite{EMSMTA} in the context of conformal wavelets. Here we only point out that, 
replacing $X=Z'^\dag Z$ in  (\ref{antikernel}) and using determinant and Wigner's $\cD$-matrix properties 
\cite{Louck3}, one easily realizes that that (\ref{antikernel}) reproduces (\ref{closure}).

Let us introduce bracket notation and put
\be
 \la {}{}_{q_a,q_b}^{j,m}|Z\ra \equiv \varphi_{q_a,q_b}^{j,m}(Z)\det(\sigma_0-Z^\dag Z)^{\lambda/2}.
 \label{Ruhlket}\ee
(We remove the label $\lambda$ from the definition of $|{}{}_{q_a,q_b}^{j,m}\rangle$ for the sake of brevity). This makes  
${\cal H}_\lambda(\mathbb D_4)$ a \emph{reproducing kernel Hilbert space}, that is, a Hilbert space of functions 
$\varphi$ in which pointwise evaluation $\varphi(Z)$ is a continuous linear functional. 
The resolution of the identity for an orthonormal basis in ${\cal H}_\lambda(\mathbb D_4)$ then adopts the form
\be 1=\sum^{\infty}_{m=0}\sum_{j\in \mathbb N/2}\sum^{j}_{q_a,q_b=-j}
|{}{}_{q_a,q_b}^{j,m}\ra \la{}{}_{q_a,q_b}^{j,m}|,
\ee
and the formal ket $|Z\ra$ is
\be
|Z\ra=\det(\sigma_0-Z^\dag Z)^{\lambda/2}\sum^{\infty}_{m=0}\sum_{j\in\mathbb N/2}\sum^{j}_{q_a,q_b=-j}\varphi_{q_a,q_b}^{j,m}(Z)
|{}{}_{q_a,q_b}^{j,m}\ra.\label{u4cs}
\ee
Actually, we can identify $|Z\ra$ with the coherent state $|\psi^\lambda_g\ra$ in \eqref{cs} up to a phase. 
From the coherent state overlap
\be
\la Z'|Z\ra=\frac{\det(\sigma_0-Z'^\dag Z')^{\lambda/2}\det(\sigma_0-Z^\dag Z)^{\lambda/2}}{\det(\sigma_0-Z'^\dag Z)^\lambda}\label{u4csov}
\ee
we see that $|Z\ra$ is normalized. Moreover, using the orthogonality properties of the homogeneous polynomials
$\varphi_{q_a,q_b}^{j,m}(Z)$, it is direct to prove the announced resolution of unity \eqref{resol}, now written as:
\be
1=c_\lambda\int_{\mathbb D_4} |Z\ra\la Z|\left.d\mu(g)\right|_{\mathbb D_4}.\label{u4csclos}
\ee
It is interesting to compare the $U(2,2)/U(2)^2$ CS  \eqref{u4cs} with the well known $U(1,1)/U(1)^2$, 
or Bargmann index-$\kappa$ (with $\kappa>1/2$), CS
\be
|z\rangle=(1-|z|^2)^{\kappa}\sum_{n=0}^\infty\varphi_n(z)|\kappa,n\rangle,\;\; 
\varphi_n(z)=\binom{2\kappa+n-1}{n}^{1/2}z^{n},\label{su2cs}
\ee
with $z\in\mathbb D_1=\{z\in\mathbb C: 1-|z|^2>0\}$ (the stereographic projection of the hyperboloid $U(1,1)/U(1)^2$ 
onto the unit disk), for which the coherent state overlap
and the resolution of the identity acquire the form
\be
\langle z'|z\rangle=\frac{(1-|z'|^2)^\kappa(1-|z|^2)^\kappa}{(1-\bar{z'}z)^{2\kappa}},\;\;
1=\frac{2\kappa-1}{\pi}\int_{{\mathbb D_1}}|z\rangle\langle z|\frac{|dz|}{(1-|z|^2)^2}.\label{overresosu2}
\ee
We perceive a similar structure between $U(2,2)/U(2)^2$ and $U(1,1)/U(1)^2$ CS, although the case  $U(2,2)/U(2)^2$ is more
involved and can be regarded as a  generalized (matrix $Z$) version of the standard (scalar $z$) case.

We finish this section with an explicit form of the unirep of $U(2,2)$ on $\mathcal{H}_\lambda(\mathbb D_4)$ 
in the form of a Corollary.

{\cor\label{cor3} For any holomorphic function $\phi\in\mathcal{H}_\lambda(\mathbb D_4)$ and any $g'\in U(2,2)$, the following
action
\be [{\cal U}^\lambda_{g'}\phi](Z)\equiv \det(D'^\dag-B'^\dag
Z)^{-\lambda}\phi(Z'), \; Z'=(A'^\dag Z-C'^\dag)(D'^\dag-B'^\dag Z)^{-1} \label{reprerest}\ee
defines a square-integrable unitary irreducible representation of $U(2,2)$ on $\mathcal{H}_\lambda(\mathbb D_4)$.
}\\
Note that if we define $\psi(g)\equiv\psi^\lambda_0(g)\phi(Z), Z=Z(g)$, then
\be
[{\cal U}^\lambda_{g'}\phi](Z)=(\psi^\lambda_0(g))^{-1}[\mathcal{U}(g')\psi](g).
\ee
The unitarity of $\mathcal{U}$ in $L^2(U(2,2),d\mu)$ directly implies the unitarity of ${\cal U}^\lambda$
in $\mathcal{H}(\mathbb D_4)$. Irreducibility follows from the fact that, for example, for $\phi(Z)=1$, the
transformed function
\be
[{\cal U}^\lambda_{g'}\phi](Z)\equiv \det(D'^\dag-B'^\dag
Z)^{\lambda}=\sum^{\infty}_{m=0}\sum_{j\in\mathbb N/2}\sum^{j}_{q_a,q_b=-j}
c_{q_a,q_b}^{j,m}(g')\varphi_{q_a,q_b}^{j,m}(Z)
\ee
is expanded in terms of all basis functions $\varphi_{q_a,q_b}^{j,m}(Z)$ with non-zero coefficients
$c_{q_a,q_b}^{j,m}(g')=\det(D'^\dag)^{-\lambda} \overline{\varphi_{q_a,q_b}^{j,m}({B'D'^{-1}})}$, as follows
from \eqref{closure}.

{\rem\label{antivac} Instead of the ground state $\psi_0^\lambda(g)=\det(D)^{-\lambda}$, we could also have chosen 
$\psi_0^\lambda(g)=\det(A)^{-\lambda}$, for which we would have arrived to a square-integrable unitary irrep of $U(2,2)$ on the 
space $\overline{\mathcal{H}_\lambda(\mathbb D_4)}$ of anti-holomorphic functions $\phi(Z^\dag)$.}

\section{Infinitesimal generators, matrix elements and operator symbols}\label{sec3} 

Let us denote by $\mathfrak{P}^\mu, \mathfrak{M}_{\mu\nu}, \mathfrak{D}$ and $\mathfrak{K}^\mu$ the infinitesimal 
(differential) generators of the 
finite action \eqref{reprerest} fulfilling the same commutation relations as the 
matrix generators  $P^\mu, M^{\mu\nu}, D$ and $K^\mu$ in \eqref{confalgamma}. 
Writting $Z=z_\mu\sigma^\mu, z_\mu\in\mathbb C$, 
$z^2=z_\mu z^\mu$, $\partial^\mu=\partial/\partial z_\mu$,  
these generators have the following expression:
\bea
\mathfrak{M}^{\mu\nu}=z^\mu \partial^\nu-z^\nu \partial^\mu, &
\mathfrak{D}=z_\mu\partial^\mu+\lambda,\nn\\
\mathfrak{P}^{\mu}=\partial^\mu, &
\mathfrak{K}^{\mu}= z^2 \mathfrak{P}^\mu-2z^\mu \mathfrak{D},\label{difop}
\eea
Let us compute their action on the orthonormal basis functions \eqref{basisfunc}. Firstly we see that the homogeneous polynomials in 
\eqref{basisfunc} are eigenfunctions of the dilation generator  $\mathfrak{D}$ 
\be
\mathfrak{D}\varphi_{q_a,q_b}^{j,m}=(2j+2m+\lambda)\varphi_{q_a,q_b}^{j,m},\label{interimbalance}
\ee
with eigenvalue $2j+2m+\lambda$, where $2j+2m$ is the homogeneity degree of $\varphi_{q_a,q_b}^{j,m}$ and $\lambda$ is the scale dimension.  
Similarly, we can compute the infinitesimal action of spacetime translations
\bea
\mathfrak{P}^0\varphi_{q_a,q_b}^{j,m}&=&
C_{q_a,q_b}^{j,m+2j+1}\varphi_{q_a-\um,q_b-\um}^{j-\um,m}+
C_{-q_a+\um,-q_b+\um}^{j+\um,m}\varphi_{q_a-\um,q_b-\um}^{j+\um,m-1}+\nn\\ &&
C_{-q_a,-q_b}^{j,m+2j+1}\varphi_{q_a+\um,q_b+\um}^{j-\um,m}+
C_{q_a+\um,q_b+\um}^{j+\um,m}\varphi_{q_a+\um,q_b+\um}^{j+\um,m-1}\,,\nn\\
\mathfrak{P}^1\varphi_{q_a,q_b}^{j,m}&=&
C_{-q_a,q_b}^{j,m+2j+1}\varphi_{q_a+\um,q_b-\um}^{j-\um,m}-
C_{q_a+\um,-q_b+\um}^{j+\um,m}\varphi_{q_a+\um,q_b-\um}^{j+\um,m-1}+\nn\\ &&
C_{q_a,-q_b}^{j,m+2j+1}\varphi_{q_a-\um,q_b+\um}^{j-\um,m}-
C_{-q_a+\um,q_b+\um}^{j+\um,m}\varphi_{q_a-\um,q_b+\um}^{j+\um,m-1}\,,\nn\\
\mathfrak{P}^2\varphi_{q_a,q_b}^{j,m}&=&
iC_{-q_a,q_b}^{j,m+2j+1}\varphi_{q_a+\um,q_b-\um}^{j-\um,m}-
iC_{q_a+\um,-q_b+\um}^{j+\um,m}\varphi_{q_a+\um,q_b-\um}^{j+\um,m-1}-\nn\\ &&
iC_{q_a,-q_b}^{j,m+2j+1}\varphi_{q_a-\um,q_b+\um}^{j-\um,m}+
iC_{-q_a+\um,q_b+\um}^{j+\um,m}\varphi_{q_a-\um,q_b+\um}^{j+\um,m-1}\,,\nn\\
\mathfrak{P}^3\varphi_{q_a,q_b}^{j,m}&=&
C_{q_a,q_b}^{j,m+2j+1}\varphi_{q_a-\um,q_b-\um}^{j-\um,m}+
C_{-q_a+\um,-q_b+\um}^{j+\um,m}\varphi_{q_a-\um,q_b-\um}^{j+\um,m-1}-\nn\\ &&
C_{-q_a,-q_b}^{j,m+2j+1}\varphi_{q_a+\um,q_b+\um}^{j-\um,m}-
C_{q_a+\um,q_b+\um}^{j+\um,m}\varphi_{q_a+\um,q_b+\um}^{j+\um,m-1}\,,\label{lowering}
\eea
and relativistic uniform accelerations
\bea
\mathfrak{K}^0\varphi_{q_a,q_b}^{j,m}&=&
-C_{q_a,q_b}^{j,m+1}\varphi_{q_a-\um,q_b-\um}^{j-\um,m+1}-
C_{-q_a,-q_b}^{j,m+1}\varphi_{q_a+\um,q_b+\um}^{j-\um,m+1}-\nn\\ &&
C_{-q_a+\um,-q_b+\um}^{j+\um,m+2j+1}\varphi_{q_a-\um,q_b-\um}^{j+\um,m}-
C_{q_a+\um,q_b+\um}^{j+\um,m+2j+1}\varphi_{q_a+\um,q_b+\um}^{j+\um,m}\,,\nn\\
\mathfrak{K}^1\varphi_{q_a,q_b}^{j,m}&=&
C_{-q_a+\um,q_b+\um}^{j+\um,m+2j+1}\varphi_{q_a-\um,q_b+\um}^{j+\um,m}+
C_{q_a+\um,-q_b+\um}^{j+\um,m+2j+1}\varphi_{q_a+\um,q_b-\um}^{j+\um,m}-\nn\\ &&
C_{q_a,-q_b}^{j,m+1}\varphi_{q_a-\um,q_b+\um}^{j-\um,m+1}-
C_{-q_a,q_b}^{j,m+1}\varphi_{q_a+\um,q_b-\um}^{j-\um,m+1}\,,
\nn\\
\mathfrak{K}^2\varphi_{q_a,q_b}^{j,m}&=&
-iC_{-q_a+\um,q_b+\um}^{j+\um,m+2j+1}\varphi_{q_a-\um,q_b+\um}^{j+\um,m}+
iC_{q_a+\um,-q_b+\um}^{j+\um,m+2j+1}\varphi_{q_a+\um,q_b-\um}^{j+\um,m}+\nn\\ &&
iC_{q_a,-q_b}^{j,m+1}\varphi_{q_a-\um,q_b+\um}^{j-\um,m+1}-
iC_{-q_a,q_b}^{j,m+1}\varphi_{q_a+\um,q_b-\um}^{j-\um,m+1}\,,
\nn\\
\mathfrak{K}^3\varphi_{q_a,q_b}^{j,m}&=&
C_{q_a+\um,q_b+\um}^{j+\um,m+2j+1}\varphi_{q_a+\um,q_b+\um}^{j+\um,m}-
C_{-q_a+\um,-q_b+\um}^{j+\um,m+2j+1}\varphi_{q_a-\um,q_b-\um}^{j+\um,m}+\nn\\ &&
C_{-q_a,-q_b}^{j,m+1}\varphi_{q_a+\um,q_b+\um}^{j-\um,m+1}-
C_{q_a,q_b}^{j,m+1}\varphi_{q_a-\um,q_b-\um}^{j-\um,m+1}\,,\label{rising}
\eea
with
\be
C_{q_a,q_b}^{j,m}\equiv\frac{\sqrt{(j+q_a)(j+q_b)m(\lambda+m-2)}}{\sqrt{2j(2j+1)}}.
\ee
The infinitesimal generators of rotations $U(2)^2=U_a(2)\times U_b(2)$ are the angular momentum operators
$\mathfrak{S}_{aj}=\um(\mathfrak{M}_{0j}-i\epsilon_{jkl}\mathfrak{M}_{kl})$
and $\mathfrak{S}_{bj}=\um(\mathfrak{M}_{0j}+i\epsilon_{jkl}\mathfrak{M}_{kl})$. The action of the angular-momentum  third component is 
\be
\mathfrak{S}_{\ell 3}\,\varphi_{q_a,q_b}^{j,m}=q_\ell\, \varphi_{q_a,q_b}^{j,m},\; \ell=a, b \label{jzeta}
\ee
and the action of the ladder  angular-momentum operators is 
\bea
\mathfrak{S}_{\ell\pm}\,\varphi_{q_a,q_b}^{j,m}=\sqrt{(j\mp q_\ell)(j\pm q_\ell+1)}\,
\varphi_{q_a\pm\delta_{\ell,a},q_b\pm\delta_{\ell,b}}^{j,m},\; \ell=a,b \label{jpm}
\eea
where $\mathfrak{S}_{a\pm}=\mathfrak{S}_{a 1}\mp i\mathfrak{S}_{a 2}$ and  $\mathfrak{S}_{b\pm}=
\mathfrak{S}_{b 1}\pm i\mathfrak{S}_{b 2}$. Note that $\mathfrak{S}_{a\pm}$ and $\mathfrak{S}_{b\pm}$
have conjugated definitions ($\pm\leftrightarrow\mp$).
This fact is related to the transformation property of wave functions in \eqref{reprerest} which,
for pure rotations ($C'=0=B', \,
A'=V_a, D'=V_b; V_\ell\in SU(2), \ell=a,b$) gives $[{\cal U}^\lambda_{g'}\phi](Z)=\phi(V_a^\dag ZV_b)$, so that rotations 
$V_a$ are represented by the inverse $V_a^\dag$. 

For completeness, we also give the action of $U(2)^2$-invariant (i.e., commuting with $\mathfrak{M}_{\mu\nu}$) 
quadratic operators: 
\bea
\mathfrak{M}_{\mu\nu} \mathfrak{M}^{\mu\nu}\varphi_{q_a,q_b}^{j,m}&=&-8j(j+1)\varphi_{q_a,q_b}^{j,m},\nn\\
\mathfrak{P}^\mu\mathfrak{P}_\mu\varphi_{q_a,q_b}^{j,m}&=& 4\sqrt{m(2j+m+1)(\lambda+m-2)(\lambda+2j+m-1)}
\varphi_{q_a,q_b}^{j,m-1},\nn\\
\mathfrak{K}^\mu\mathfrak{K}_\mu\varphi_{q_a,q_b}^{j,m}&=& 4\sqrt{(m+1)(2j+m+2)(\lambda+m-1)(\lambda+2j+m)}
\varphi_{q_a,q_b}^{j,m+1},\nn\\
\mathfrak{K}^\mu\mathfrak{P}_\mu\varphi_{q_a,q_b}^{j,m}&=& -4(2j^2+m(m+\lambda-2)+j(2m+\lambda-1))
\varphi_{q_a,q_b}^{j,m},\nn\\
\mathfrak{P}^\mu\mathfrak{K}_\mu\varphi_{q_a,q_b}^{j,m}&=& -4(2j^2+(m+2)(m+\lambda)+j(2m+\lambda+3))
\varphi_{q_a,q_b}^{j,m}.\label{ppkk}
\eea
One can verify that $\mathfrak{P}^\mu\mathfrak{K}_\mu+\mathfrak{K}^\mu\mathfrak{P}_\mu=
2\mathfrak{P}^\mu\mathfrak{K}_\mu+8\mathfrak{D}$, as deduced from the original commutation relation $\lb K_\mu,P_\nu\rb = 2(\eta_{\mu\nu}
D+M_{\mu\nu})$. 
With these ingredients, the value of the quadratic Casimir operator 
\be \mathfrak{C}_2= \mathfrak{D}^2-\um  \mathfrak{M}_{\mu\nu} \mathfrak{M}^{\mu\nu}+
\um( \mathfrak{P}_\mu  \mathfrak{K}^\mu+ \mathfrak{K}_\mu  \mathfrak{P}^\mu)\label{Casimir}\ee
in the Hilbert space $\mathcal{H}_\lambda(\mathbb D_4)$ is easily computed and gives:
\be
\mathfrak{C}_2\varphi_{q_a,q_b}^{j,m}=\lambda(\lambda-4)\varphi_{q_a,q_b}^{j,m},
\;\;\forall j,m,q_a,q_b.\label{casieigen}
\ee
We shall also provide the operator symbols (the expectation value in $|Z\ra$) of the previous operators. Given the 
differential representation \eqref{difop} of any operator $\mathfrak{O}$ we can simply compute its symbol as
\be
\la \mathfrak{O}\ra\equiv \la Z|\mathfrak{O}|Z\ra=\frac{\mathfrak{O}\det(\sigma^0-Z^\dag Z)^{-\lambda}}{ 
\det(\sigma^0-Z^\dag Z)^{-\lambda}}.
\ee
With this information, the operator symbols of \eqref{difop} and their quadratic scalar combinations \eqref{ppkk} are

\bea
&\la\mathfrak{D}\ra= \lambda\frac{1-\det(Z^\dag Z)}{\det(\sigma^0-Z^\dag Z)},\; 
\la\mathfrak{P}^{\mu}\ra= 2\lambda\frac{\bar z_\mu-\det(Z^\dag Z)z^\mu}{\det(\sigma^0-Z^\dag Z)},&\\
&\la \mathfrak{K}^{\mu}\ra=
\det(Z^\dag Z)\la\mathfrak{P}^{\mu}\ra-2z^\mu\la\mathfrak{D}\ra,\; 
\la \mathfrak{M}^{\mu\nu}\ra= 
z^\mu \la\mathfrak{P}^{\nu}\ra-z^\nu \la\mathfrak{P}^{\mu}\ra,&\nn\\
& \la\mathfrak{D}^2\ra=\frac{\lambda+1}{\lambda}\la\mathfrak{D}\ra^2-
\lambda\frac{1+\det(Z^\dag Z)}{\det(\sigma^0-Z^\dag Z)},\; 
\la\mathfrak{P}^\mu\mathfrak{P}_\mu\ra=\frac{4\lambda(\lambda-1)\det(Z^\dag)}{\det(\sigma^0-Z^\dag Z)},\;
\la\mathfrak{K}^\mu\mathfrak{K}_\mu\ra=\frac{4\lambda(\lambda-1)\det(Z)}{\det(\sigma^0-Z^\dag Z)},&\nn\\ 
&\la\mathfrak{P}^\mu\mathfrak{K}_\mu\ra=2\left(\lambda 
\frac{\lambda-3+\tr(Z^\dag Z)+(1+\lambda)\det(Z^\dag Z)}{\det(\sigma^0-Z^\dag Z)}-
\frac{\lambda+1}{\lambda}\la\mathfrak{D}\ra^2\right),& \nn\\  
&\la\mathfrak{K}^\mu\mathfrak{P}_\mu\ra=\la\mathfrak{P}^\mu\mathfrak{K}_\mu\ra+8\la\mathfrak{D}\ra,\; 
\la\mathfrak{M}_{\mu\nu} \mathfrak{M}^{\mu\nu}\ra=2\la\mathfrak{D}^2\ra+\la\mathfrak{K}^\mu\mathfrak{P}_\mu\ra+
\la\mathfrak{P}^\mu\mathfrak{K}_\mu\ra-2\lambda(\lambda-4),&\nn
\eea
where the last one is a consequence of \eqref{Casimir} and \eqref{casieigen}. Actually, one can verify that the star commutator of symbols 
$[\la\mathfrak{O}_1\ra,\la\mathfrak{O}_2\ra]_*\equiv\la\mathfrak{O}_1\ra*\la\mathfrak{O}_2\ra-\la\mathfrak{O}_2\ra*\la\mathfrak{O}_1\ra$, with 
star product $\la\mathfrak{O}_1\ra*\la\mathfrak{O}_2\ra\equiv\la \mathfrak{O}_1\mathfrak{O}_2\ra$, defines a representation 
$[\la\mathfrak{O}_1\ra,\la\mathfrak{O}_2\ra]_*=\la[\mathfrak{O}_1,\mathfrak{O}_2]\ra$ of the $su(2,2)$ Lie algebra.

\section{Oscillator realization, massive compounds and excitons}\label{oscisec}

It is well known the oscillator (Jordan-Schwinger) realization of the $SU(1,1)$ generators $\mathcal{Q}_3, \mathcal{Q}_\pm$ in terms
of two bosonic modes $a$ and $b$ (see \cite{bosonsu2su11} for a general discussion on boson realizations of $su(1,1)$ and $su(2)$) as
\be
\mathcal{Q}_3=\um(a^\dag a+b^\dag b+1),\; \mathcal{Q}_+=a^\dag b^\dag,\; \mathcal{Q}_-=ab,\label{schwingersu2}
\ee
and the expression of Bargmann index-$\kappa$ basis states $|\kappa,n\ra, \, n=0,1,2,\dots,\infty$, in terms of Fock states
($|0\ra$ denotes the Fock vacuum)
\be
|n_a\ra\otimes |n_b\ra=\frac{(a^\dag)^{n_a}(b^\dag)^{n_b}}{\sqrt{n_a!n_b!}}|0\ra\label{Focku2}
\ee
as
\be
|\kappa,n\ra=\frac{(a^\dag)^{n}(b^\dag)^{n+2\kappa-1}}{\sqrt{n!(n+2\kappa-1)!}}|0\ra=
\frac{\varphi_n(a^\dag)}{\sqrt{\frac{(2\kappa+n-1)!}{(2\kappa-1)!}}}
\frac{\varphi_{2\kappa+n-1}(b^\dag)}{\sqrt{\frac{(4\kappa+n-2)!}{(2\kappa-1)!}}}|0\ra=
|n\ra_a\otimes |n+2\kappa-1\ra_b,\label{basisinfocksu2}
\ee
where we have used the monomials 
$\varphi_n$ in \eqref{su2cs} as operator functions, since this notation will be generalized in a natural way 
later in eq. \eqref{basisinfock2} for a Fock representation of the basis functions $|{}{}_{q_a,q_b}^{j,m}\ra$ of 
${\cal H}_\lambda(\mathbb D_4)$. Note that there is always an excess of $n_b-n_a=2\kappa-1$ $b$-type   
over $a$-type  quanta, which leads to the constraint $b^\dag b-a^\dag a=2\kappa-1$. 
The lowest weight state $|\kappa,0\ra=\frac{(b^\dag)^{2\kappa-1}}{\sqrt{(2\kappa-1)!}}|0\ra$ can be 
regarded as a  boson condensate of $2\kappa-1$ $b$-type  particles, and the rest of states $|\kappa,n\ra$ as pair $ab$  excitations 
(``excitons'') above this condensate. The $SU(1,1)$  CS \eqref{su2cs} can also be written as
\be
|z\ra=
{(1-|z|^2)^{\kappa}}{e^{z \mathcal{Q}_+}}|\kappa,0\rangle 
.\label{su2csboson}
\ee
Is is interesting to see that, defining $\mathcal Z=\begin{pmatrix} a^\dag\\ b \end{pmatrix}$ and
$\mathcal Z^\dag=\begin{pmatrix} a & b^\dag \end{pmatrix}$, the $U(1,1)$ generators
\eqref{schwingersu2} can be compactly written as
\be
\mathcal{Q}_\mu=\um \mathcal Z^\dag \sigma_\mu\Gamma \mathcal Z, \label{QZ} \ee
with $\sigma_\mu=\eta_{\mu\nu}\sigma^\nu$, $\Gamma=\mathrm{diag}(-1,1)$, 
$\mathcal{Q}_\pm=-i\mathcal{Q}_2\pm \mathcal{Q}_1$ and the extra generator
$\mathcal{Q}_0=\um(b^\dag b-aa^\dag) =\kappa-1$ (linear Casimir) is related to the excess of $b$- over $a$-type quanta. 
Later in eq. \eqref{PL} we shall relate this excess of quanta with the helicity $s=\kappa-1/2$ 
(see also Section \ref{comments} for a Lagrangian  interpretation inside a 
twistor description of massless conformal particles). Note that for $\kappa\geq 1/2$ 
we have only positive values of the helicity $s$. Negative values of $s$ come from the alternative choice 
$|\kappa,0\ra_-=\frac{(a^\dag)^{2\kappa-1}}{\sqrt{(2\kappa-1)!}}|0\ra$ for the lowest weight state, now 
regarded as a  boson condensate of $2\kappa-1$ $a$-type  particles. Actually, the quotient $U(1,1)/U(1)^2$ is the two-sheet 
hyperboloid leading to two orbits $\mathbb D_1^\pm$ related to positive and negative helicity $s$.

The natural (minimal) generalization of this $U(1,1)$ construction to $U(2,2)$ requires four bosonic modes $a_1, a_2, b_1, b_2$, for which Fock states are
\be
|n_{a}^1\ra\otimes |n_{a}^2\ra\otimes |n_{b}^1\ra\otimes |n_{b}^2\ra=
\frac{(a_1^\dag)^{n_a^1}(a_2^\dag)^{n_a^2}(b_1^\dag)^{n_b^1}(b_2^\dag)^{n_b^2}}{\sqrt{n_a^1!n_a^2!n_b^1!n_b^2!}}|0\ra, 
\label{symmetricbasis}
\ee
with $n_a^j, n_b^j\in\mathbb N$ the corresponding occupation numbers. Defining now 
\be\mathcal Z^\dag=(a_1, a_2, b_1^\dag, b_2^\dag),\label{calzetasym}
\ee
the Jordan-Schwinger realization of the sixteen $u(2,2)$ generators in \eqref{confalgamma}, 
compactly written as $X_{\mu\nu}=\{D,M_{\mu\nu},P_\mu,K_\nu,I_4\}$ (we are 
adding the $4\times 4$ identity matrix $X_{00}=I_4$), is given by
\be
\mathcal{X}_{\mu\nu}=\mathcal{Z}^\dag X_{\mu\nu}\Gamma \mathcal Z,\label{bosrepresym} \ee
where now $\Gamma=\mathrm{diag}(-1,-1,1,1)$. 
Indeed, one can easily verify that $[\mathcal{X}_{\mu\nu},\mathcal{X}_{\mu'\nu'}]=\mathcal Z^\dag
[X_{\mu\nu},X_{\mu'\nu'}]\Gamma \mathcal Z$, and therefore \eqref{bosrepresym} defines a
(unitary) representation of $u(2,2)$ in the Fock space \eqref{symmetricbasis}. Fixing again the excess of 
$b$- over $a$-type quanta as $n_b^1+n_b^2-n_a^1-n_a^2=2\kappa-1$ [which means to fix the linear Casimir 
$\mathcal{X}_{00}=2\kappa-3$, or the helicity $\mathcal S=\um\mathcal{X}_{00}+1$], 
the basis states for fixed $\kappa$ can now be labeled in terms of three non-negative integers $\vec{n}=(n_1,n_2,n_3)$ as
\be
|\kappa,\vec{n}\ra=\frac{(a_1^\dag)^{n_1}(a_2^\dag)^{n_2}(b_1^\dag)^{n_3}(b_2^\dag)^{2\kappa-1+n_1+n_2+n_3}}
{\sqrt{n_1!n_2!n_3!(2\kappa-1+n_1+n_2+n_3)!}}|0\ra,\;\; n_1,n_2,n_3=0,1,\dots,\infty
\label{basisinfocksu22sym}
\ee
This corresponds to the so called \emph{ladder representations} of $U(2,2)$ describing massless particles with 
helicity $s=\kappa-\um$ (see 
\cite{Mack-Todorov} for other basis and the irreducibility of this representation when restricted 
to the Poincar\'e subgroup). 
Indeed, if we define $\mathcal{P}^\mu$, $\mathcal{K}^\mu$ and 
 $\mathcal{M}^{\mu\nu}$  the boson realization \eqref{bosrepresym} of  four-momentum $P^\mu$,  `four-acceleration' $K^\mu$ 
and Lorentz matrix ${M}^{\mu\nu}$ in \eqref{confalgamma}, then one can verify that 
$\mathcal{P}^\mu\mathcal{P}_{\mu}=0=\mathcal{K}^\mu\mathcal{K}_{\mu}$ (zero mass) and 
\be
\mathcal{W}^\mu|\kappa,\vec{n}\ra=(\kappa-\um)\mathcal{P}^\mu|\kappa,\vec{n}\ra,\; \forall \vec{n}\in\mathbb{N}^3,\; 
\mu=0,1,2,3,\label{PL}
\ee
with $\mathcal{W}_\alpha=\frac{i}{2}\epsilon_{\alpha\mu\nu\beta}\mathcal{M}^{\mu\nu}\mathcal{P}^\beta$ the Pauli-Lubanski operator and 
$s=\kappa-\um$ the helicity.

This (Jordan-Schwinger) oscillator realization is in fact extensible to general $U(p,q)$ \cite{Todorov2,couplingUpq} and it became popular after \cite{u66},  
who discussed the use of $U(6,6)$ to classify hadrons; in this case barions and antibarions belong to mutually conjugate representations 
with respect to $U(6)$. 

This (discrete, most degenerate) representation of $U(2,2)$ is related
to the quotient $\mathbb D_3=U(2,2)/[U(2,1)\times U(1)]$ (the pseudo complex projective space $\mathbb CP^3$) whose points
$\vec{z}=(z_1,z_2,z_3)\in \mathbb C^3$ (in a certain patch) label the CS given by the expansion 
\be
|\vec{z}\ra=(1+|z_1|^2-|z_2|^2-|z_3|^2)^{\kappa}\sum_{n=0}^\infty\sum_{m=0}^n\sum_{l=0}^m 
\varphi_{n}^{ml}(\vec{z})|\kappa,(n-m,m-l,l)\ra,\label{bosoncsu4rpoj}
\ee
in terms of the basis functions \eqref{basisinfocksu22sym}, where the coefficients 
\be
\varphi_{n}^{ml}(\vec{z})=\sqrt{\binom{2\kappa+n-1}{n}\binom{n}{m}\binom{m}{l}}
(-1)^{\frac{n-m}{2}}z_1^{n-m}z_2^{m-l}z_3^l,
\ee
are homogeneous polynomials of degree $n$ in three complex variables $\vec{z}$. 
The values of $z_i$ are not arbitrary but must fulfill 
$1+|z_1|^2-|z_2|^2-|z_3|^2>0$ in the present patch. These CS are normalized and 
also verify a resolution of the identity similar to the one in \eqref{overresosu2} but replacing the
$\mathbb D_1$ integration measure by the corresponding $\mathbb D_3$ integration measure. The CS \eqref{bosoncsu4rpoj} 
can also be written as the exponential action [equivalent of 
\eqref{su2csboson}] of pair creation operators $Q_{+1}=a_1^\dag b_2^\dag, Q_{+2}=a_2^\dag b_2^\dag, Q_{+3}=b_1^\dag b_2^\dag$ on the 
ground state $|\kappa,\vec{0}\ra=\frac{(b_2^\dag)^{2\kappa-1}}{\sqrt{(2\kappa-1)!}}|0\ra$ as
\be
|\vec{z}\ra=(1+|z_1|^2-|z_2|^2-|z_3|^2)^{\kappa}e^{iz_1 Q_{+1}+z_2 Q_{+2}+z_3Q_{+3}}|\kappa,\vec{0}\ra.\label{expoCSCP3}
\ee
As already discussed for 
$U(1,1)$, again we have two conformal orbits $\mathbb D_3^\pm$ corresponding to phase-spaces of negative and 
positive helicity zero-mass particles. The ground state for negative helicity particles can be namely taken as  
$|\kappa,\vec{0}\ra_-=\frac{(a_1^\dag)^{2\kappa-1}}{\sqrt{(2\kappa-1)!}}|0\ra$, where there is an excess of 
$2\kappa-1$ $a$-type over $b$-type quanta.

However, these are still not the CS \eqref{u4cs} we are dealing with in this article. Actually, the CS \eqref{u4cs}
will be related to massive conformal particles (see later in Sec. \ref{comments} for a Lagrangian picture). The question is:
is there a boson realization like \eqref{bosoncsu4rpoj} and \eqref{expoCSCP3} but for the CS
\eqref{u4cs} labeled by points $Z$ in the Cartan domain $\mathbb D_4$?. The answer is positive and 
it will be given later in Proposition \ref{propcsboson2}. Indeed, there is another way of extending the $U(1,1)$ 
construction to  $U(2,2)$ by defining now
\be
\mathcal Z=\begin{pmatrix}
            \mathbf a^\dag\\ \mathbf b
           \end{pmatrix}=
\begin{pmatrix}
\begin{matrix} a_0^\dag & a_2^\dag\\ a_1^\dag & a_3^\dag
\end{matrix}
\\ \begin{matrix} b_0 & b_1\\ b_2 & b_3
\end{matrix} \end{pmatrix},\label{calzeta}\ee
which can be seen as a ``compound''  $\mathcal Z=(\mathcal Z_1,\mathcal Z_2)$ of two zero-mass systems 
$\mathcal Z_1=(a_0^\dag,a_1^\dag,b_0,b_2)^t$ and $\mathcal Z_2=(a_2^\dag,a_3^\dag,b_1,b_3)^t$ with certain constraints 
given below (see later in this Section for an explicit proof). The new oscillator realization 
\be
\mathcal{X}_{\mu\nu}=\tr({\mathcal Z}^\dag X_{\mu\nu}\Gamma \mathcal Z),\label{bosrepre} \ee
of the sixteen $u(2,2)$ matrix generators $X_{\mu\nu}$ in \eqref{confalgamma} [plus identity $X_{00}$],  defines a
(unitary) representation of $u(2,2)$ in the Fock space with basis states
\be
|\mathbf n_a\ra \otimes |{\mathbf n}_b\ra=\left|\begin{matrix} n_a^0 & n_a^1\\ n_a^2 & n_a^3
\end{matrix}\right>\otimes \left|\begin{matrix} n_b^0 & n_b^1\\ n_b^2 & n_b^3
\end{matrix}\right>=\prod_{\mu=0}^3
\frac{(a^\dag_\mu)^{n_a^\mu}(b^\dag_\mu)^{n_b^\mu}}{\sqrt{n_a^\mu!n_b^\mu!}}|0\ra.\label{grassmannbasis2}
\ee
Let us look for the expression of the basis states $|{}{}_{q_a,q_b}^{j,m}\ra$ in \eqref{Ruhlket} in terms of the
Fock basis \eqref{grassmannbasis2}. The $U(2,2)$ analogue of the $U(1,1)$ (linear Casimir) constraint 
$2\mathcal{Q}_0=b^\dag b-aa^\dag=2(\kappa-1)$  [remember \eqref{QZ}] on the basis states $|\kappa,n\rangle$ \eqref{basisinfocksu2},  
here adopts the matrix form 
\be\mathcal Z^\dag \Gamma\mathcal Z=\mathbf b^\dag \mathbf b-\mathbf a \mathbf a^\dag=(\lambda-4)
\mathcal{I}_{2}\label{constraintZ}\ee
on the basis states $|{}{}_{q_a,q_b}^{j,m}\ra$, where $\mathcal{I}_{2}$
denotes the $2\times 2$ identity operator. 
In particular, there is an excess of $\sum_{\mu=0}^3 n_b^\mu-n_a^\mu=2(\lambda-2)$ of $b$- over $a$-type quanta, that is,
the linear Casimir operator $\mathcal{X}_{00}=\sum_{\mu=0}^3 b^{\dag}_\mu b_\mu-a_\mu a^{\dag}_\mu$
is fixed to $2(\lambda-4)$. From \eqref{interimbalance}, we also see that the dilation operator
$\mathcal{D}=2+\um\sum_{\mu=0}^3 (a^{\dag}_\mu a_\mu+b^{\dag}_\mu b_\mu)$ provides the relation
\be 2+\um\sum_{\mu=0}^3 (n_a^\mu+n_b^\mu)=2j+2m+\lambda,\label{eigenD2}\ee with $2j+2m$ the homogeneity degree of
$\varphi^{j,m}_{q_a,q_b}$, which also coincides with the total number of pair $ab$ excitations (``excitons'')  over 
the lowest-weight (zero homogeneity degree) ground state  $|\varphi_0\ra\equiv|{}{}_{q_a=0,q_b=0}^{j=0,m=0}\ra$, which is made 
of $2(\lambda-2)$ $b$-type quanta and can expressed in terms
of Fock states as:
\be
|\varphi_0\ra=\frac{\det(\mathbf b^\dag)^{\lambda-2}}{(\lambda-2)!\sqrt{\lambda-1}}|0\ra= \left|\begin{matrix} 0 & 0\\ 0 & 0
\end{matrix}\right>_a\otimes \sum_{k=0}^{\lambda-2}\frac{ (-1)^k}{\sqrt{\lambda-1}}\left|\begin{matrix}
\lambda-2-k & k\\ k & \lambda-2-k
\end{matrix}\right>_b.\label{lowestweight}\ee
Indeed, one can easily check that $|\varphi_0\ra$ fulfills the
constraints $\mathcal Z^\dag \Gamma\mathcal Z=(\lambda-4)
\mathcal{I}_{2}$. In order to obtain the expression of the rest of basis states $|{}{}_{q_a,q_b}^{j,m}\ra$
in terms of Fock states \eqref{grassmannbasis2}, we have firstly made use of the differential 
representation $\mathfrak{X}_{\mu\nu}$ of the bosonic operators $\mathcal{X}_{\mu\nu}$ and  
applying, step by step\footnote{We do not present here the (rather 
cumbersome) steps to get this result. We must acknowledge the benefits of  \emph{Mathematica} add-on packages like 
``Quantum Algebra'' to check this and some other 
expressions along this Section. These packages are available at \cite{Quantum}.}, 
ladder operators (\ref{lowering},\ref{rising},\ref{jpm}) and \eqref{ppkk} to the lowest-weight state 
\eqref{lowestweight}, we have finally arrived to the expression

\be
|{}{}_{q_a,q_b}^{j,m}\ra=\frac{1}{\sqrt{2j+1}}\sum_{q=-j}^{j}
\frac{\varphi^{j,m}_{q_a,q}(\mathbf{a}^\dag)}{\sqrt{\frac{(\lambda-2)!(\lambda-1)!}{(\lambda+2j+m-1)!(\lambda+m-2)!}}}
\frac{\varphi^{j,\lambda+m-2}_{q,q_b}(\mathbf{b}^\dag)}{\sqrt{\frac{(\lambda-2)!(\lambda-1)!}{(2\lambda+2j+m-3)!(2\lambda+m-4)!}}}
\;|0\ra,
\label{basisinfock2}
\ee
where we are now treating the homogeneous
polynomials $\varphi^{j,m}_{q,q'}$ in \eqref{basisfunc} as operator functions, since there is not
ordering problem (all $a_\mu^\dag$ and $b_\mu^\dag$ commute). This is the $SU(2,2)$ version of eq. \eqref{basisinfocksu2} 
for the Bargmann index-$\kappa$ basis states $|\kappa,n\ra$ of $SU(1,1)$,
with the role of $\kappa$ played now by $\lambda$ and the role of the monomials $\varphi_n(z)$ played now by the 
homogeneous polynomials $\varphi_{q_a,q_b}^{j,m}(Z)$.

In the process we have found extra restrictions to the
number $n_a^\mu$ and $n_b^\mu$ of  $a$- and $b$-type bosons like:
\be
(n_b^0+n_b^2)-(n_a^0+n_a^1)=\lambda-2=(n_b^1+n_b^3)-(n_a^2+n_a^3)\,,\label{helycoupled}
\ee
and 
\bea
n_a^0-n_a^1+n_a^2-n_a^3&=&2q_a\,,\nn\\
n_b^0+n_b^1-n_b^2-n_b^3&=&2q_b\,.\label{angular3}
\eea
The restrictions \eqref{angular3} say that the third angular momentum components $q_a$ and $q_b$, 
measure the imbalance between 
$\mu=\{0,2\}$ (spin up) and $\mu=\{1,3\}$ (spin down) $a$-type bosons and 
$\mu=\{0,1\}$ (spin up) and $\mu=\{2,3\}$ (spin down) $b$-type bosons.

The restriction \eqref{helycoupled} could be interpreted by saying that both zero-mass particles, $\mathcal{Z}_1$ and $\mathcal{Z}_2$ 
forming the compound $\mathcal{Z}=(\mathcal{Z}_1,\mathcal{Z}_2)$, carry helicity $s=(\lambda-2)/2$. Indeed, 
if we define the four-momentum $\mathcal{P}^\mu_p$,  
Lorentz $\mathcal{M}^{\mu\nu}_p$ and Pauli-Lubanski $\mathcal{W}^\mu_p$ operators for each separated 
particle $\mathcal{Z}_p$ [$\mathcal Z_1=(a_0^\dag,a_1^\dag,b_0,b_2)^t$ and $\mathcal Z_2=(a_2^\dag,a_3^\dag,b_1,b_3)^t$] of the compound 
$\mathcal{Z}=(\mathcal{Z}_1,\mathcal{Z}_2)$ as
\be
\mathcal{P}^\mu_p=\mathcal{Z}_p^\dag P^\mu \Gamma\mathcal{Z}_p,\;\; 
\mathcal{M}^{\mu\nu}_p=\mathcal{Z}_p^\dag M^{\mu\nu} \Gamma\mathcal{Z}_p,\; p=1,2,\;\; 
\mathcal{W}_\alpha^p=\frac{i}{2}\epsilon_{\alpha\mu\nu\beta}\mathcal{M}^{\mu\nu}_p\mathcal{P}^\beta_p,\; p=1,2,
\ee
then one can check, for each individual particle $p=1,2$, that
\be
\mathcal{P}^\mu_p\mathcal{P}_{p\mu}=0, \; 
\mathcal{W}^\mu_p|{}{}_{q_a,q_b}^{j,m}\ra=\frac{\lambda-2}{2}\mathcal{P}^\mu_p|{}{}_{q_a,q_b}^{j,m}\ra,\; p=1,2,
\ee
are valid for any basis vector $|{}{}_{q_a,q_b}^{j,m}\ra$, which means that both particles are massless and carry 
helicity $s_p=(\lambda-2)/2, p=1,2$. However, the compound $\mathcal{Z}=(\mathcal{Z}_1,\mathcal{Z}_2)$ 
displays a continuum mass spectrum since  
\be 
\mathcal{P}^\mu\mathcal{P}_\mu|{}{}_{q_a,q_b}^{j,m}\ra= 4\sqrt{m(2j+m+1)(\lambda+m-2)(\lambda+2j+m-1)}
|{}{}_{q_a,q_b}^{j,m-1}\ra,
\ee
with $\mathcal{P}^\mu=\tr({\mathcal Z}^\dag P^\mu\Gamma \mathcal Z)$ the compound's four-momentum, and the corresponding Pauli-Lubanski operator 
is zero $\mathcal{W}^\mu|{}{}_{q_a,q_b}^{j,m}\ra=0$, since the compound $(\mathcal{Z}_1,\mathcal{Z}_2)$ is spin-less (see e.g. \cite{spinning} 
for more general, spinning, representations of the conformal group in a geometrical setting).

The existence of a continuum mass spectrum was already adverted for the differential representation in \eqref{ppkk}. 
The fact that two massless 
particles can form a massive compound can result somehow awkward. In the Standard Model, particles acquire 
mass from the Higgs boson trough the ``spontaneous breakdown'' of the gauge symmetry. However, here we see that 
coupling massless particles can also result in a massive compound. This seems to be a profound result that deserves 
more attention and will be studied elsewhere.

We shall now study the quantum statistics of the compound under the exchange 
$(\mathcal{Z}_1,\mathcal{Z}_2)\to (\mathcal{Z}_2,\mathcal{Z}_1)$ 
of its two massless constituents for a given  $\lambda$. The result is given in the next Theorem.

{\thm\label{Statistics} Wave functions for the compound $\mathcal{Z}=(\mathcal{Z}_1,\mathcal{Z}_2)$ are symmetric under the exchange 
$(\mathcal{Z}_1,\mathcal{Z}_2)\to (\mathcal{Z}_2,\mathcal{Z}_1)$ for $\lambda$ even and antisymmetric for $\lambda$ odd. In particular, 
the basis functions \eqref{basisinfock2} verify 
\be\widetilde{|{}{}_{q_a,q_b}^{j,m}\ra}=(-1)^\lambda\, |{}{}_{q_a,q_b}^{j,m}\ra,\label{statistics}\ee
where $\widetilde{|{}{}_{q_a,q_b}^{j,m}\ra}$ is constructed as in \eqref{basisinfock2} but replacing 
\[\mathbf{a}=\begin{pmatrix} a_0 & a_1\\ a_2 & a_3 \end{pmatrix}\to \; \widetilde{\mathbf{a}}=\begin{pmatrix} a_2 & a_3\\ a_0 & a_1 \end{pmatrix},\;
\mathbf{b}=\begin{pmatrix} b_0 & b_1\\ b_2 & b_3 \end{pmatrix}\to \; \widetilde{\mathbf{b}}=\begin{pmatrix} b_1 & b_0\\ b_3 & b_2 \end{pmatrix},\]
that is, exchange of rows in $a$-type particles and exchange of columns in $b$-type particles.}

\noindent\textbf{Proof:} The proof is simple when one realizes that 
the operator functions \eqref{basisfunc} verify 
\[\varphi^{j,m}_{q_a,q_b}(\widetilde{\mathbf{b}}^\dag)=(-1)^{m}\varphi^{j,m}_{-q_a,q_b}(\mathbf{b}^\dag),\;\; 
\varphi^{j,m}_{q_a,q_b}(\widetilde{\mathbf{a}}^\dag)=(-1)^{m}\varphi^{j,m}_{q_a,-q_b}(\mathbf{a}^\dag).\] Taking into account 
that $(-1)^{2q}=(-1)^{2j}$ for any $q=-j,\dots,j$ and doing some algebraic manipulations, 
one arrives to the identity \eqref{statistics}.$\blacksquare$ 

This is a consequence of the indistinguishability of the 
two  zero-mass particles, $\mathcal{Z}_1$ and $\mathcal{Z}_2$, forming the compound 
$\mathcal{Z}=(\mathcal{Z}_1,\mathcal{Z}_2)$. The even/odd character of $\lambda$ (related to the 
helicity $(\lambda-2)/2$ of the massless particles and to the excess number $2(\lambda-2)$ of unpaired particles) 
determines the bosonic/fermionic character of the compound.

Note that the two massless particles of the compound 
are correlated (constrained) and identical. These correlations are worth studying more carefully and will be 
left for future work \cite{Entangph}.

Before finishing this Section,  we shall provide a boson realization
like \eqref{su2csboson} but for the CS
\eqref{u4cs} labeled by points $Z$ in the complex domain $\mathbb D_4$.

{\prop\label{propcsboson2} Let us denote by ${\mathcal{K}}\equiv
\mathcal{K}^{\mu}{\sigma}_\mu=-2{\mathbf{a}}^\dag{\mathbf{b}}^\dag$, with $\mathcal{K}^{\mu}=\tr({\mathcal Z}^\dag K^\mu\Gamma \mathcal Z)$ 
the oscillator realization  of the matrix $K^\mu$ in \eqref{confalgamma}. The
CS $|Z\ra$ in  \eqref{u4cs} can be written as the exponential action of creation particle-hole operators $\mathcal{K}$ on the lowest-weight 
state $|\varphi_0\ra$ as
\be
|Z\ra={\det(\sigma^0-Z^\dag Z)^{\lambda/2}}{e^{-\um \tr({Z}^t{\mathcal{K}})}}|\varphi_0\ra.\label{u4cs2}
\ee
}

\noindent\textbf{Proof:} Proving \eqref{u4cs2} is equivalent to prove that
\be
e^{-\mathcal{A}}|\varphi_0\ra=\sum^{\infty}_{m=0}\sum_{j\in\mathbb N/2}
\sum^{j}_{q_a,q_b=-j}\varphi_{q_a,q_b}^{j,m}(Z)
|{}{}_{q_a,q_b}^{j,m}\ra,\label{u4cs3}
\ee
with 
\be
\mathcal{A}\equiv \um \tr({Z}^t{\mathcal{K}})=(-1)^\mu z_\mu \mathcal{K}^\mu, \; (\mathrm{sum \ on}\, \mu).
\ee
Note  that the equivalence of the expressions \eqref{u4cs} and \eqref{u4cs2}
for CS on $U(2,2)/U(2)^2$ is the matrix counterpart of the (easier to prove) equivalence of \eqref{su2cs} and \eqref{su2csboson} for
CS on $U(1,1)/U(1)^2$. To prove \eqref{u4cs3} we shall proceed by induction on the homogeneity degree in $Z$. More precisely, 
we shall firstly prove that the identity 
\be
\frac{(-\mathcal{A})^n}{n!}|\varphi_0\rangle=\sum_{j=\frac{\mathrm{odd}(n)}{2}}^{n-\frac{\mathrm{odd}(n)}{2}}\sum_{q_a=-j}^j\sum_{q_b=-j}^j
\varphi_{q_a,q_b}^{j,\frac{n}{2}-j}(Z)|{}{}_{q_a,q_b}^{j,\frac{n}{2}-j}\ra,\label{hipo}
\ee
is true for any $n\in \mathbb N$, where  ${\mathrm{odd}(n)}=(1-(-1)^{n})/2$. In fact, it is trivially fulfilled for $n=0$. 
For $n=1$ we have that 
\bea
-\mathcal{A}|\varphi_0\ra&=&-\sqrt{\lambda}((z_3-z_0)|{}{}_{\frac{-1}{2},\frac{-1}{2}}^{\;\um,\;0}\ra-
(z_3+z_0)|{}{}_{\um,\um}^{\um,0}\ra-(z_1-iz_2)|{}{}_{\um,\frac{-1}{2}}^{\um,\;0}\ra\nonumber\\ & &
-(z_1+iz_2)|{}{}_{\frac{-1}{2},\um}^{\; \um,0}\ra)= \sum^{1/2}_{q_a,q_b=-1/2}\varphi_{q_a,q_b}^{\um,0}(Z)
|{}{}_{q_a,q_b}^{\um,0}\ra,
\eea
where we have made use of the differential representation $\mathfrak{K}^\mu$ of $\mathcal{K}^\mu$ 
in \eqref{rising} and the definition of $\varphi_{q_a,q_b}^{j,m}(Z)$ in \eqref{basisfunc}. Therefore, the hypothesis  
\eqref{hipo} is true for $n=1$. Assuming \eqref{hipo} to be true for a given $n$, and using again \eqref{rising} and 
the definition \eqref{basisfunc}, we finally arrive to
\bea
\frac{(-\mathcal{A})^{n+1}}{(n+1)!}|\varphi_0\rangle&=&\frac{-1}{n+1}\sum_{j=\frac{\mathrm{odd}(n)}{2}}^{n-\frac{\mathrm{odd}(n)}{2}}\sum_{q_a=-j}^j\sum_{q_b=-j}^j
\varphi_{q_a,q_b}^{j,\frac{n}{2}-j}(Z)\mathcal{A}|{}{}_{q_a,q_b}^{j,\frac{n}{2}-j}\ra\nonumber\\
&=&\sum_{j=\frac{\mathrm{odd}(n+1)}{2}}^{n+1-\frac{\mathrm{odd}(n+1)}{2}}\sum_{q_a=-j}^j\sum_{q_b=-j}^j
\varphi_{q_a,q_b}^{j,\frac{n+1}{2}-j}(Z)|{}{}_{q_a,q_b}^{j,\frac{n+1}{2}-j}\ra,
\eea
which states that the hypothesis \eqref{hipo} is true for $n+1$, thus completing the proof by induction. The expansion 
of the exponential $e^{-\mathcal{A}}=\sum_{n=0}^\infty\frac{(-\mathcal{A})^n}{n!}$ completes the proof of \eqref{u4cs3} 
and then of \eqref{u4cs2}.$\blacksquare$

The operator ${\mathcal{K}}^\mu=-\tr({\mathbf{a}}^\dag{\mathbf{b}}^\dag\sigma_\mu)$ creates particle-hole pairs (excitons) and 
 ${\mathcal{P}}^\mu=-\tr({\mathbf{b}}{\mathbf{a}}\sigma^\mu)$ annihilates excitons. It is said in the literature that excitons 
 are almost bosons \cite{excitonsknox} since exciton creation and annihilation operators fulfill bosonic commutation relations 
 plus corrections in the number of particle-hole pairs which arise from the interaction between excitons. In our case, if we 
 renormalize the two-body creation  ${\mathcal{K}}_\mu/\sqrt{2(\lambda-2)}\equiv {\mathcal{E}}_\mu^\dag$ and annihilation 
 ${\mathcal{P}}_\mu/\sqrt{2(\lambda-2)}\equiv {\mathcal{E}}_\mu$ operators with the square root of the number $2(\lambda-2)$ of 
 unpaired particles, and we use the fact that $|{}{}_{q_a,q_b}^{j,m}\ra$ are eigenstates of $\mathcal{D}$ with eigenvalue $n_e+\lambda$ 
 [$n_e=2j+2m$ is the number of excitons; remember (\ref{interimbalance},\ref{eigenD2})], then the  basic 
commutator $\lb {\mathcal{K}}_\mu,{\mathcal{P}}_\nu\rb = 2(\eta_{\mu\nu}
{\mathcal{D}}+{\mathcal{M}}_{\mu\nu})$ says that (we restrict ourselves to $\mu=\nu$ for simplicity)
\be
\la {}{}_{q_a,q_b}^{j,m}|\lb {\mathcal{E}}_\mu^\dag,{\mathcal{E}}_\mu\rb|{}{}_{q_a,q_b}^{j,m}\ra=2\eta_{\mu\mu}
\frac{\lambda+n_e}{2(\lambda-2)}\simeq \eta_{\mu\mu}\left(1+ O(\frac{n_e}{\lambda})\right),\ee
for a large number of unpaired particles, $\lambda\gg 1$. This result agrees with the general fact that excitons are almost bosons 
as long as $n_e\ll \lambda$ [note that the commutator for the temporal $\mu=0$ component has the reversed desired sign].

The expression \eqref{u4cs2} shows the coherent state $|Z\ra$ as a ``Bose-Einstein condensate'' of excitons. 
Interesting physical phenomena of Bose-Einstein condensation of excitons and biexcitons can be found in 
 \cite{excitonBEC}. We believe that our abstract construction of coherent states of excitons can provide 
 an interesting framework to study  physical applications in this context.

\section{Lagrangian picture and physical interpretations\label{comments}}

Let us propose a physical interpretation of the previous abstract mathematical construction by making use 
of the \emph{twistor particle picture} (see e.g. \cite{Penrose1,Penrose2,Penrose-MacCallum,Penrose3,Penrose-Rindler}) and $U(2,2)$ nonlinear sigma model Lagrangians 
(see e.g. \cite{spinning,JGP,MacfarlanePLB}). General integrable Hamiltonian systems on $SU(2,2)$ and $SU(p,q)$ have also been 
discussed in \cite{Olmo1,Olmo2}.

The twistor space $\mathbb T$ is a 4-dimensional complex vector space $\mathbb C^4$ with points 
$\zeta=(\alpha_1,\alpha_2,\beta_1,\beta_2)^t\in\mathbb C^4$ 
($t$ means transpose), and their duals 
$\zeta^\dag=(\bar\alpha_1,\bar\alpha_2,\bar\beta_1,\bar\beta_2)$. 
The bilinear product (related to the helicity) $\varsigma=\zeta^\dag\Gamma\zeta=|\beta_1|^2+|\beta_2|^2-|\alpha_1|^2-|\alpha_2|^2$ 
is invariant under the natural linear action of $U(2,2)$ in \eqref{su22}. It measures the imbalance between, let us 
say, ``particle'' ($\beta$ spinor) and ``antiparticle'' ($\alpha$ spinor), which transform under  conjugated $U(2)$ transformations,  
$U(2)_\alpha\times U(2)_\beta\subset U(2,2)$. It is well known that twistors describe massless particles. 
The phase space of a massless conformal particle is the pseudo complex projective space $\mathbb D_3=U(2,2)/[U(2,1)\times U(1)]$, which  
carries two open orbits $\mathbb D_3^\pm$ corresponding to the sign of the helicity. We shall take 
representatives $\zeta\in \mathbb D_3^\pm$ normalized according to  $\varsigma=\zeta^\dag\Gamma\zeta=\pm 1$. Note that the first two 
columns of $g\in U(2,2)$ in eq. \eqref{su22} belong to $\mathbb D_3^-$ and the last two ones belong to $\mathbb D_3^+$. 
A Lagrangian whose quantization leads to the oscillator realization discussed in \eqref{bosrepresym} and 
\eqref{basisinfocksu22sym} is formulated in terms of a nonlinear $\sigma$-model as follows. Although 
we are in a quantum mechanical setting (finite number of particles), we shall consider the general many particle 
case (field theory) to which the former case reduces for zero spatial dimensions. Therefore, consider the twistor $\zeta$ as a 
field on spacetime $\zeta(x^\mu)$ [for a single particle, we would consider a function of just time $\zeta(x^0)$] and denote 
$\partial_\mu\zeta=\partial\zeta/\partial x^\mu$  [resp.  $\dot\zeta=d\zeta/dx^0$ for mechanics]. In $\mathbb D_3$, 
the twistors $\zeta$ and $\zeta u$, with $u\in U(1)$ an arbitrary phase, are equivalent. Therefore, 
$\zeta$ transforms as $\zeta\to g\zeta u$ under global $g\in U(2,2)$ and local $u\in U(1)$. The Lagrangian 
$L=\um \partial_\mu\zeta^\dag\Gamma\partial^\mu\zeta$ is not $U(1)$ gauge invariant, but the minimal 
coupling $\partial_\mu\to D_\mu-A_\mu$, with $U(1)$-gauge field $A_\mu=\varsigma\zeta^\dag\Gamma\partial_\mu\zeta$, renders the 
Lagrangian
\[L_\varsigma=\um \partial_\mu\zeta^\dag(1-\varsigma\Gamma\zeta\zeta^\dag)\Gamma\partial^\mu\zeta\] 
$U(1)$-gauge invariant in each orbit $\mathbb D_3^\varsigma$ with $\varsigma=\pm 1$ and $(1-\varsigma\Gamma\zeta\zeta^\dag)\Gamma$ 
a projector. Other gauge invariant terms linear in ``velocity'' can be added to the Lagrangian in the 
quantum mechanical case \cite{spinning}. In the quantization process, the negative-energy problem is overcome by assigning $\zeta\to\mathcal Z$, with 
$\mathcal{Z}$ the operator-valued twistor in \eqref{calzetasym}; that is, we assign annihilation operators $(b_1,b_2)$ to particles $\beta$ and 
creation  operators  $(a_1^\dag,a_2^\dag)$ to antiparticles $\alpha$. This reminds the negative-energy problem 
for Dirac particles, although the solution there entails the imposition of anticommutation relations. 
The quantum mechanical helicity $\mathcal S=\um\mathcal{Z}^\dag\Gamma\mathcal{Z}+1$ can take any half-integer value.

Let us now discuss the more involved massive case. A single massive particle turns out to be described in terms of two 
(or more) twistors $\zeta_1,\zeta_2\in\mathbb C^4$ arranged as:
\be
\mathbf{Z}=(\zeta_1,\zeta_2)=
\begin{pmatrix}
\begin{matrix} \alpha_{11} & \alpha_{12}\\ \alpha_{21} & \alpha_{22}
\end{matrix}
\\ \begin{matrix} \beta_{11} & \beta_{12}\\ \beta_{21} & \beta_{22}
\end{matrix} \end{pmatrix}\label{Zbf}
\ee
The (massless) constraint  $\varsigma=\zeta^\dag\Gamma\zeta=\pm 1$ now adopts the form $\varSigma=\mathbf{Z}^\dag\Gamma\mathbf{Z}=\pm I_2$ 
(with $I_2$ the $2\times 2$ identity matrix). The phase space of a massive conformal particle is the complex pseudo-Grassmann manifold 
$\mathbb D_4=U(2,2)/[U(2)\times U(2)]$ of 2-planes in $\mathbb C^4$, which  
carries two open orbits $\mathbb D_4^\pm$ corresponding to the sign $\varsigma=\pm 1$ of $\varSigma=\varsigma I_2$. 
Note that the first block-(two)-column of $g\in U(2,2)$ in eq. \eqref{su22} belongs to $\mathbb D_4^-$ and the last 
one belongs to $\mathbb D_4^+$. In $\mathbb D_4$, 
the bi-twistors $\mathbf{Z}$ and $\mathbf{Z} U$, with $U\in U(2)$ an arbitrary rotation, are equivalent. Therefore, 
$\mathbf{Z}$ transforms as $\mathbf{Z}\to g\mathbf{Z} U$ under global $g\in U(2,2)$ and local $U\in U(2)$. 
Minimal coupling $\partial_\mu\to D_\mu-A_\mu$, with $U(2)$-gauge field $A_\mu=\varsigma\mathbf{Z}^\dag\Gamma\partial_\mu\mathbf{Z}$, 
renders the Lagrangian
\[L_\varsigma=\um \tr[\partial_\mu\mathbf{Z}^\dag(1-\varsigma\Gamma\mathbf{Z}\mathbf{Z}^\dag)\Gamma\partial^\mu\mathbf{Z}]\] 
$U(2)$-gauge invariant in each orbit $\mathbb D_4^\varsigma$, with $(1-\varsigma\Gamma\mathbf{Z}\mathbf{Z}^\dag)\Gamma$ 
a projector. One can eliminate four complex fields (out of the original eight) from the theory by making a choice of gauge
\be
\mathbf{Z}(Z)=\begin{pmatrix} Z\Delta_2\\ \Delta_2
           \end{pmatrix},
\ee
which leaves four complex degrees of freedom $Z$ parametrizing a point on $\mathbb D_4$ in \eqref{cartandomain}, 
with $\Delta_2$ defined in \eqref{blockortho} (note that this choice corresponds to $\varsigma=1$). This factorization 
has to do with the Iwasawa decomposition \eqref{Iwasawa}. 

In the quantization process, the negative-energy problem is now overcome by assigning $\mathbf{Z}\to\mathcal Z$, with 
$\mathcal{Z}$ the operator-valued bi-twistor in \eqref{calzeta}; that is,
\be\begin{pmatrix} \alpha_{11} & \alpha_{12}\\ \alpha_{21} & \alpha_{22}\end{pmatrix}\to 
\begin{pmatrix} a_0^\dag & a_2^\dag\\ a_1^\dag & a_3^\dag
\end{pmatrix},\;  \begin{pmatrix} \beta_{11} & \beta_{12}\\ \beta_{21} & \beta_{22}
\end{pmatrix}\to  \begin{pmatrix} b_0 & b_1\\ b_2 & b_3
\end{pmatrix}.\ee
The lowest weight state $|\varphi_0\ra$ in \eqref{lowestweight} can be 
regarded as a  boson condensate of $2(\lambda-2)$ $b$-type  quanta (let's say, ``matter'' or ``particles''), and the rest of states 
$|{}{}_{q_a,q_b}^{j,m}\ra$ as pair $ab$  (``matter-antimatter'' or ``particle-hole'') excitations (``excitons'', in condensed matter jargon) 
above this condensate. Another possibility consists of considering an excess of $a$-type quanta  
over $b$-type quanta. This particle excess (or, more precisely, $\lambda$) determines the statistics 
(either bosonic or fermionic) of the physical states under the interchange of the two particles of the compound [interchange 
of columns in \eqref{calzeta} or in \eqref{Zbf}] according to \eqref{statistics}. One would be tempted to 
relate this asymmetry with the parity violation in weak nuclear forces or, even further, with the matter-antimatter 
asymmetry in the universe; but, for the time being, these are just simple speculations.

We believe that our construction will be useful not only for the better understanding of the structure of 
conformal quantum particles, but also for general pairing systems like nuclear pairing phenomenon or 
superconductivity in solid state physics. The quantum number $\lambda$ in these models would be related to 
the so called ``seniority number'', which counts the number of unpaired nucleons or electrons in each case. We 
must say that the infinite-dimensional character of the $U(2,2)$ [or general $U(N,N)$] bosonic representation 
should be in principle more appropriate to describe superconductivity phenomena, where the number of 
free electrons is very large, although projection techniques on finite number of particles could also be 
applied. In fact, from the oscillator realization of (a certain number of copies of) $su(2)$ and $su(1,1)$, 
many interesting pairing many-body Hamiltonians have been built like Bardeen-Cooper-Schrieffer, 
Lipkin-Meshkov-Glick, etc (see e.g. \cite{dukelskirmp04,dukelskinp05} in the context of Richardson-Gaudin models). The construction 
made here for $U(2,2)$ can be easily generalized to $U(N,N)$ following the guidelines of the extended MacMahon-Schwinger master theorem 
proved by us in \cite{EMSMTA}; this also opens the possibility of dealing with even more complex situations (more degrees of freedom).

The Euclidean [$\Gamma=\mathrm{diag}(1,1,1,1)\to U(4)$] version of the present construction has been 
recently studied in \cite{GrassmannU4}, where a physical interpretation inside the bilayer fractional quantum Hall effect 
has been put forward. In particular, $a$ and $b$ correspond to the top and bottom layers (the pseudospin) and 
the index $\lambda$ is related to the number of flux quanta bound to a bi-fermion 
in the \emph{composite fermion} picture of Jain for fractions of the filling factor $\nu=2$. We recommend the 
interested reader to have a look at this Euclidean $U(4)$ version to perceive similarities and differences with the 
present $U(2,2)$ construction.

\section*{Acknowledgements}

Work partially supported by the
Spanish MICINN, University of Granada and Junta de Andaluc\'\i a under projects FIS2011-29813-C02-01, 
PP2012-PI04 and FQM1861 respectively.

\end{document}